# Organic Solvent Based Synthesis of Gold Nanoparticle - Semiconducting 2H-MoS$_2$ Hybrid Nanosheets

*Abhijit Ganguly,[†] Olga Trovato,[#] Shanmughasundaram Duraisamy,[†] John Benson,[‡] Yisong Han,[†] Cristina Satriano[#] and Pagona Papakonstantinou[\*,†]*

[†]School of Engineering, Engineering Research Institute, Ulster University, Newtownabbey BT37 0QB, United Kingdom

[#] Department of Chemical Sciences, University of Catania, Viale Andrea Doria 6, 95125 Catania, Italy

[‡]2-DTech, Core Technology Facility, 46 Grafton St., Manchester M13 9NT, United Kingdom

[\*]**Corresponding Author.** E-mail address: p.papakonstantinou@ulster.ac.uk

**ABSTRACT:**

The development of simple, versatile strategies for the synthesis of gold nanoparticles (AuNPs) on semiconducting transition metal dichalcogenides (TMDC) layers is of increasing scientific and technological interest in photocatalysis, optical sensing, and optoelectronics sectors, but challenges exist on the nucleation and hybridization of AuNPs with the TMDC basal plane. At present, the widely used aqueous solution approaches suffer from poor dispersion of produced hybrids as well as from limited growth and coverage of the AuNPs on the TMDC semiconducting plane, since Au nanoclusters nucleate preferentially at the electron rich defect edges, which act as reducing agents and not on the defect free basal plane. Here, we report for the first time, the controlled synthesis of AuNPs on the basal plane of semiconducting molybdenum disulfide nanosheets (2H-$MoS_2$ NSs) via a N,N-dimethylformamide (DMF)-based hot-injection synthesis route. This organic solvent-based synthesis route eliminates problems of poor dispersion of AuNPs@2H-$MoS_2$ NS hybrids, whereas at the same time maintains the semiconducting crystalline quality of the pristine 2H-$MoS_2$ NSs. In addition, the study establishes the important role of trisodium citrate, on enhancing the nucleation and improving the hybridization of AuNPs on 2H-$MoS_2$ NSs as evidenced by the induced p-type doping. This organic solvent synthesis approach can be adopted for other hybrid systems opening the way for controlled hybridization of semiconducting layers with metal nanoparticles.

- **INTRODUCTION**

Hybrid nanostructures, consisting of two-dimensional (2D) layers decorated with nanoparticles, have been attracting a great deal of interest in the nanomaterial-research community as building blocks for functional devices and systems for many applications. With the recent intense interest in 2D layer-structured transition-metal dichalcogenides (TMDC), like molybdenum disulfide ($MoS_2$), for their utilization in photonics, sensors and catalysis sectors, the investigation of noble-metal decorated TMDC hybrids has attracted increasing attention as an ideal strategy to modify their electronic structure and tune their performance.[1-11] Furthermore, the hybrid nanostructure provides an attractive multifunctional platform for combining the individual properties derived from both the metal and the 2D material, with a potential for triggering new phenomena.

So far, a number of strategies have been employed for the synthesis of hybrids consisting of $MoS_2$ NSs decorated with gold nanoparticles (AuNPs@$MoS_2$). The simplest approach[3-4, 9, 11] of physically mixing AuNPs and $MoS_2$ NSs, has the advantage of utilizing pre-selected AuNPs with definite sizes and shapes. However, in this strategy, chemical bonding of AuNPs to the underlying $MoS_2$, which plays a critical role on device performance for most applications, is limited. The most effective tactic for achieving intimate bonding at the $MoS_2$–Au interface lies in the $Au^0$-nucleation and growth of AuNPs directly on $MoS_2$ NS, via reduction of the Au-precursors.[6-14]

In this regard, the approach, which exploits the redox chemistry of $MoS_2$ and Au-precursor for producing hybrids in aqueous medium, has received a great deal of attention recently.[2, 5, 9-11] The redox reaction between electron rich $MoS_2$ defect sites and $Au^+$ ions leads to the spontaneous nucleation of $Au^0$. Notably, in the majority of published works, gold nucleation was reported for MoS$_2$ NSs of metallic phase (M-MoS$_2$), produced by Li intercalation, where dominance of defects on the basal plane promoted the MoS$_2$/$Au^+$ redox reaction.[2-3] Nevertheless, many applications such as photocatalysis, optical sensing, and optoelectronics, favor the presence of a semiconducting-TMDC/metallic-NP interface to increase absorption, enhance photogeneration rate and/or achieve light induced charge separation.[15] As a result, the growth of AuNPs on semiconducting 2H-MoS$_2$ is been sought. However, whereas the anchoring of AuNPs on M-MoS$_2$ has been explored thoroughly,[2-3] strategies for synthesis of AuNPs on semiconducting 2H phase TMDCs (2H-TMDC), (AuNPs@2H-TMDC), are currently at incipient stages.[10, 16] To tackle this challenge, routes for producing solvent-dispersible hybrids, with a controlled nucleation and coverage of AuNPs on the 2H-TMDC basal plane are being avidly sought.

Recently, the functionalization of 2H-TMDC, using chemical reactions of gold precursors in aqueous solution, has been explored.[16] However, the Au-nucleation was limited mainly at the electron rich edges,[16] due to the absence of highly energetic defects on the basal plane, whereas the efficient synthesis of hybrids was hindered by the poor dispersion of semiconducting 2H-MoS$_2$ NSs in water. Lately, we reported the production of 2H-MoS$_2$ NSs, via room temperature ionic liquid (RTIL) assisted grinding method combined with sequential centrifugation steps.[17] Such mechanically exfoliated 2H-MoS$_2$ NSs retain their crystalline quality after the RTIL-assisted grinding, where the RTIL not only acts as a lubricant but also protects the sheets, inhibiting possible oxidation of 2H-MoS$_2$ during exfoliation. As a result, the redox chemistry of $Au^+$ ions with these almost-defect-free and poorly dispersible in water, mechanically exfoliated 2H-MoS$_2$ NSs, is expected to be highly ineffective.

It is well known, that semiconducting 2H-MoS$_2$ NSs are well dispersed in organic solvents like N,N-dimethylformamide (DMF).[17-18] Moreover, pioneering studies[19-20] have successfully demonstrated that DMF is able to act as a mild reducing agent for chloroauric acid $HAuCl_4$, forming gold atomic clusters, when heated at a desired temperature. Importantly, the reducing ability of DMF can be increased substantially at higher temperatures, rendering the reaction temperature as the most critical parameter in the DMF-based Au-synthesis route. It was found by Liu et al.[20] that the hydrolysis of tetrachloroaurate ion ($AuCl_4^-$) by DMF becomes most effective at a temperature of 140 ºC. However, reports on DMF-based synthesis of AuNPs are very limited, because of the low reduction efficiency of the common gold precursors in DMF,[21] compared to water. Moreover, it should be noted that in the case of AuNPs@2H-TMDC hybrid synthesis, the parallel synthesis of "free" AuNPs originating from inadequate hybridization of AuNPs to 2D semiconducting supports (2H-TMDC NSs) is a typical problem.

In this work for the first time, we demonstrate the tailored nucleation of AuNPs on the basal plane of 2H-$MoS_2$ NSs, through well controlled chemical conditions, employing a DMF-based hot-injection synthesis route.[19-20] A number of advantages have been demonstrated. Firstly, compared to previous approaches utilizing aqueous solutions,[2,5] this method exhibits advantages of controlled nucleation and growth of AuNPs on the basal plane of 2H-$MoS_2$ NSs. Secondly, the semiconducting crystalline quality of mechanically exfoliated 2H-$MoS_2$ is preserved after hybridization with AuNPs. Thirdly, our study establishes the crucial role of sodium citrate ($Na_3Ct$)[22-23] on controlling the NP size and distribution on semiconducting 2H-$MoS_2$ NSs as well as on increasing the efficiency of the hybridization of AuNPs on 2H-$MoS_2$ surfaces. The hybridization is evidenced from p-type doping of 2H-$MoS_2$ NSs by Au;[1, 3-4, 8, 24-25] it greatly improves the interfacial charge transport between the semiconducting nanosheets and enhances the electrocatalytic efficiency for hydrogen evolution reaction (HER) compared to pristine 2H-$MoS_2$ NSs. The proposed synthesis route can be adopted for the controlled fabrication of other hybrid structures comprised of metal NPs on 2D layered supports, which are dispersed in organic solvents.

• EXPERIMENTAL METHODS

Here, the synthesis of semiconducting 2H-MoS$_2$ nanosheets, via an ionic liquid assisted exfoliation method, and their AuNPs@2H-MoS$_2$ hybrids, via a N,N-dimethylformamide (DMF)-based hot-injection chemical route, are described. Additional experimental details, regarding chemicals, materials and characterization methods are presented in the *Supporting Information* (*SI*).

**Synthesis of semiconducting 2H-MoS$_2$ Nanosheets.** Here, the semiconducting 2H-MoS$_2$ NSs were synthesized by grinding high purity bulk MoS$_2$ powder via ionic liquid assisted exfoliation method as reported in our earlier publication.[17] Briefly, the process involved the mechanical grinding of MoS$_2$ platelets in an adequate quantity of room temperature ionic liquid (RTIL) coupled with sequential centrifugation size selection steps. During grinding, the RTIL protected every newly exposed 2H-MoS$_2$ surface by adsorbing onto the surface, keeping the sheets separated and avoiding restacking. After grinding for sufficiently long duration, the resulting gel was subjected to multiple washing steps in a mixture of DMF and acetone, in a gradually increasing ratio, to remove the RTIL. Finally, the clean ground product, consisting of an assortment of sheets of various sizes and thicknesses, was dispersed homogeneously in pure DMF and was subjected to sequential centrifugation steps with increasing centrifugation speed from 500 to 10,000 rpm, as described earlier.[17] The sequential centrifugation of the supernatant at progressively higher centrifugation speeds for longer durations allowed the isolation of thin and small particles. Large and thick platelets were pelleted at low speeds and small durations with comparatively higher yield. For our investigation on the synthesis of AuNPs@2H-MoS$_2$ hybrids, the 2H-MoS$_2$ NSs, pelleted after the centrifugation at 1000 rpm (abbreviated as *M*), were chosen due to the substantial higher yield, relative to the thinnest NSs pelleted from the centrifugation at 10,000 rpm.

**Synthesis of AuNPs@2H-MoS$_2$ Hybrid Nanosheets.** *Hot-Injection DMF-based Synthesis of AuNPs on 2H-MoS$_2$ Nanosheets.* The basic steps of our approach for the hot-injection chemical synthesis of AuNPs on 2H-MoS$_2$ NSs in DMF medium are illustrated in Figure 1. Mimicking the approach proposed by Kawasaki et al.,[19] firstly, the well-dispersed 2H-MoS$_2$ NSs ($M_{MoS_2}$ = 5mg) in DMF solution ($V_{DMF}$ = 10 ml) are heated to 140 ºC, under vigorous stirring in a reflux system (Step#1, Fig. 1). After achieving 140 ºC, at Step#2 (Fig. 1), a freshly prepared aqueous solution of gold precursor (HAuCl$_4$) is injected directly into the hot & stirred MoS$_2$-DMF solution ($t_{HAuCl_4}$ = 140 ºC). The final concentration of HAuCl$_4$ is maintained at 1mM ($C_{HAuCl_4}$ = 1 mM).

The final solution is kept at 140 ºC under vigorous stirring for another 30 mins ($t_{Rxn}$ = 30 min), after the addition of HAuCl$_4$. This AuNPs@2H-MoS$_2$ hybrid is labeled as *D* (please see the list of AuNPs@2H-MoS$_2$ hybrids, Table at the bottom of Fig. 1, also Table S1 in the *Supporting Information* (*SI*) for further details). Purposely, the final reaction time is kept shorter, in order to evaluate the $Au^0$-nucleation stage. Growth of AuNPs, for longer reaction time, would suppress the chances to study the nucleation stage and hence the MoS$_2$-Au interaction stage.

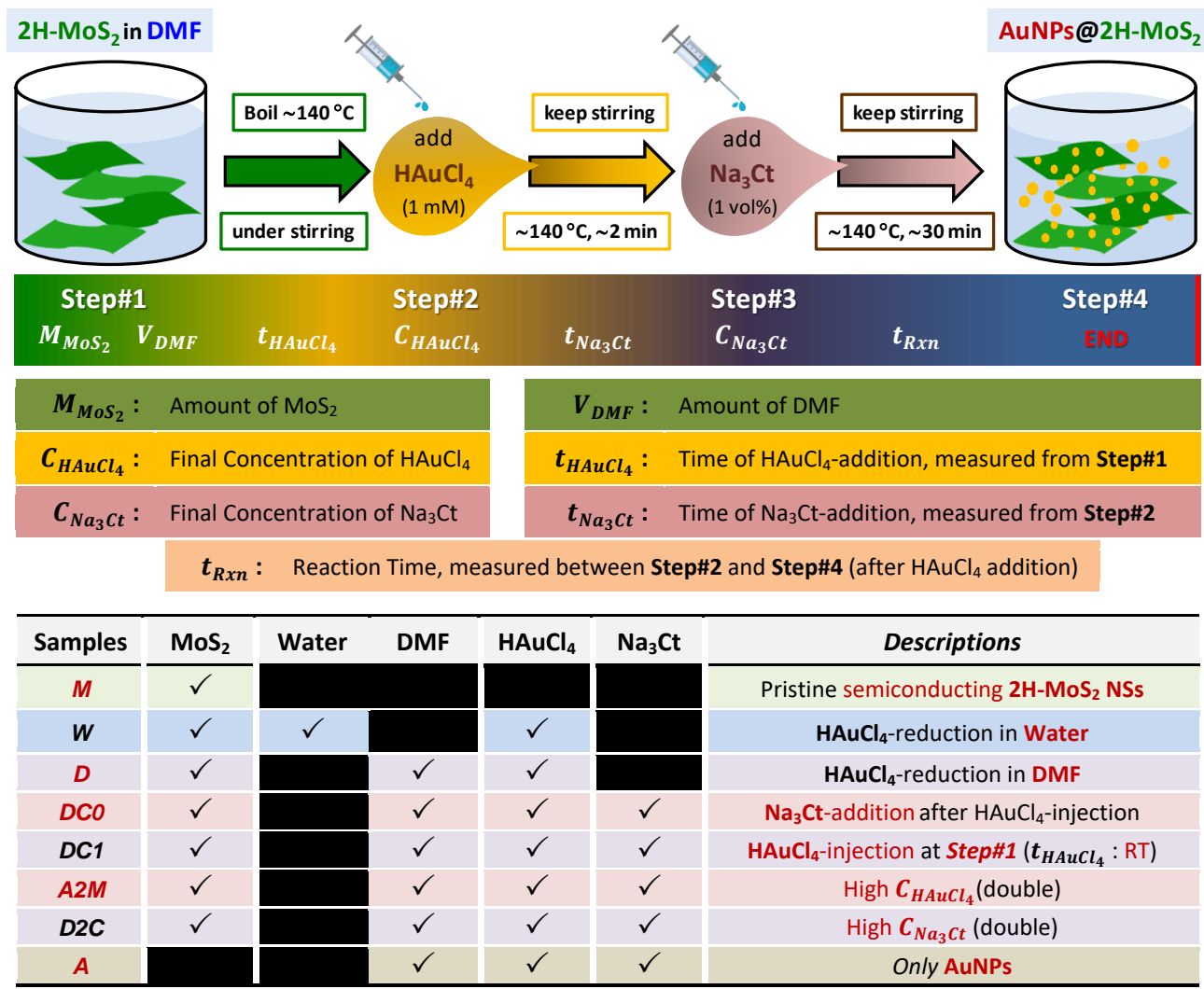

**Figure 1.** The top schematic illustrates the basic steps of hot-injection chemical synthesis of gold nanoparticles (AuNPs) on semiconducting 2H-MoS$_2$ nanosheets (NSs) hybrids, (AuNPs@2H-MoS$_2$). The bottom table provides the list of AuNPs@2H-MoS$_2$ hybrids together with their synthesis parameters; further details are provided in the *Supporting Information* (*SI*), Table S1.

For all the synthesized AuNPs@2H-MoS$_2$ hybrids, following cooling of the solution to room temperature, the product is collected and purified by centrifugation at 10k rpm for 1 hr. Subsequently, the yellowish supernatant is separated out, and the sediment is collected and dispersed in fresh DMF under adequate ultrasonication (or mechanical vibration). Repeating this purification step, finally a clear supernatant is observed and the precipitated sediment is collected and dispersed in fresh DMF. At the final stage, the solution is subjected to mild centrifugation of around 2k rpm for 1 hr, in order to separate and collect the heavier AuNPs@2H-MoS$_2$ hybrid nanosheets, which are subsequently dried, weighed and labeled.

***Effect of Solvent: DMF vs. Water, (D vs. W).*** In order to find the effect of solvent, similar route is also performed by replacing DMF with water. In this case, the final temperature was maintained at 100 °C ($t_{HAuCl_4}$) instead of 140 °C; keeping all other steps same. Here, the final product is labeled as *W*.

***Effect of Na$_3$Ct: Synthesis with Na$_3$Ct, (DC0).*** For the 2$^{nd}$ phase of our study, we employ the Na$_3$Ct, as a secondary reducing and stabilizing agent. Following the initial steps as described above, at Step#3 (Fig. 1), a freshly prepared aqueous solution of Na$_3$Ct is injected quickly ($t_{Na_3Ct}$ ≈ 1-2 min) after the addition of HAuCl$_4$ solution (Step#2). The final concentration of Na$_3$Ct is maintained at $C_{Na_3Ct}$ = 1 vol%, following the well-established and well-practiced Na$_3$Ct reduction protocol of gold colloidal solution.[22-23] The solution is kept at 140 °C under vigorous stirring for another ~30 mins ($t_{Rxn}$). The product is labeled as *DC0*.

***Effect of $t_{HAuCl_4}$ : Time of HAuCl$_4$-injection, (DC1).*** For the AuNPs@2H-MoS$_2$ hybrid labeled as *DC1*, adapting the synthesis approach used by Liu et al.,[20] HAuCl$_4$ solution is added to the MoS$_2$-DMF solution at the beginning (at Step#1, $t_{HAuCl_4}$ = at RT). Subsequently, the whole mixture is heated to 140 °C, under vigorous stirring in a reflux system. After reaching the final temperature of 140 °C, a freshly prepared aqueous solution of Na$_3$Ct is added quickly, following the recipe of *DC0* ($t_{Na_3Ct}$ =140 °C).

**Effect of HAuCl$_4$:MoS$_2$ Ratio, (A2M).** For comparison, the HAuCl$_4$:MoS$_2$ ratio is increased by doubling the $C_{HAuCl_4}$ to 2 mM, (the product labeled as *A2M)*, keeping all other steps same as for *DC0*.

**Effect of $C_{Na_3Ct}$ : Final Concentration of Na$_3$Ct, (D2C).** The mass-amount of Na$_3$Ct is doubled, $C_{Na_3Ct}$ = 2 vol%, maintaining all other steps same as *DC0;* the product labeled as *D2C*.

- **RESULTS AND DISCUSSION**

Here, we present the successful demonstration of citrate, Na$_3$Ct, modified DMF-based hot-injection chemical synthesis of AuNPs@2H-MoS$_2$ hybrids. Through a systematic investigation of synthesis parameters, such as injection time ($t_{HAuCl_4}$), concentration ($C_{HAuCl_4}$) of gold precursor HAuCl$_4$, and concentration of citrate Na$_3$Ct ($C_{Na_3Ct}$), a mechanistic understanding on the role of these parameters in the DMF-based HAuCl$_4$-reduction route has been obtained. The hybridization of AuNPs with 2H-MoS$_2$ NSs was confirmed via p-type doping. The advantage of Na$_3$Ct, as a secondary reducing and stabilizing agent, was evident from morphological studies and improved electrocatalytic response to HER.

**Hot-Injection DMF-based Synthesis of AuNPs on 2H-MoS$_2$ Nanosheets.** In the first demonstration of DMF-based hot-injection chemical synthesis of AuNPs, by Liu et al.,[20] the gold precursor (HAuCl$_4$) and DMF mixture was heated to the desired temperature (under continuous stirring) to form gold atomic clusters. Later, Kawasaki et al.[19] improved the process by injecting HAuCl$_4$ into hot DMF solution (at the desired temperature) achieving homogeneous reduction, hence avoiding the formation of bulk metals byproducts.

The main steps of our hot-injection chemical synthesis strategy for the production of hybrids (AuNPs@2H-MoS$_2$ NS) in DMF medium are based on Kawasaki et al.'s protocol[19] and are illustrated in Figure 1. Firstly, well-dispersed 2H-MoS$_2$ NSs in DMF solution (abbreviated as *M*) were heated to 140 ºC, under rigorous stirring in a reflux system (Step#1, Fig. 1). When the desired temperature of 140 ºC was reached, a freshly prepared aqueous solution of gold precursor (HAuCl$_4$) was injected directly into the hot and stirred MoS$_2$-DMF solution (depicted as Step#2). The hybrid product of this synthesis-strategy is labeled as *D* (Table in Fig. 1, for further details please see Table S1 in *SI*).

Intentionally, the reaction time after the addition of HAuCl$_4$ was kept short ($t_{Rxn} \approx 30$ min), in order to evaluate the $Au^+$-nucleation stage and initial AuNP growth. Morphological studies, using SEM, demonstrate clearly the successful AuNP formation on semiconducting 2H-MoS$_2$ NSs (Fig. 2). Backscattered electron images (BEI) are also presented since they are particularly useful on providing a clear visualization of AuNPs on the 2H-MoS$_2$ NSs.

*D* hybrids exhibit a scattered distribution ($N_D$: particle number density $\approx 1.03$ particles/µm$^2$) and formation of large Au particles (mean diameter, $D_m > 293$ nm, with large standard deviation of NP-size, $SD > \pm 154$ nm) due to the aggregation of AuNPs on 2H-MoS$_2$ NSs (Fig. 2a). Further evidence is provided by TEM images, which reveal the presence of micron-sized Au particles, suffering from serious agglomeration (*D*, Figs. 3a-b).

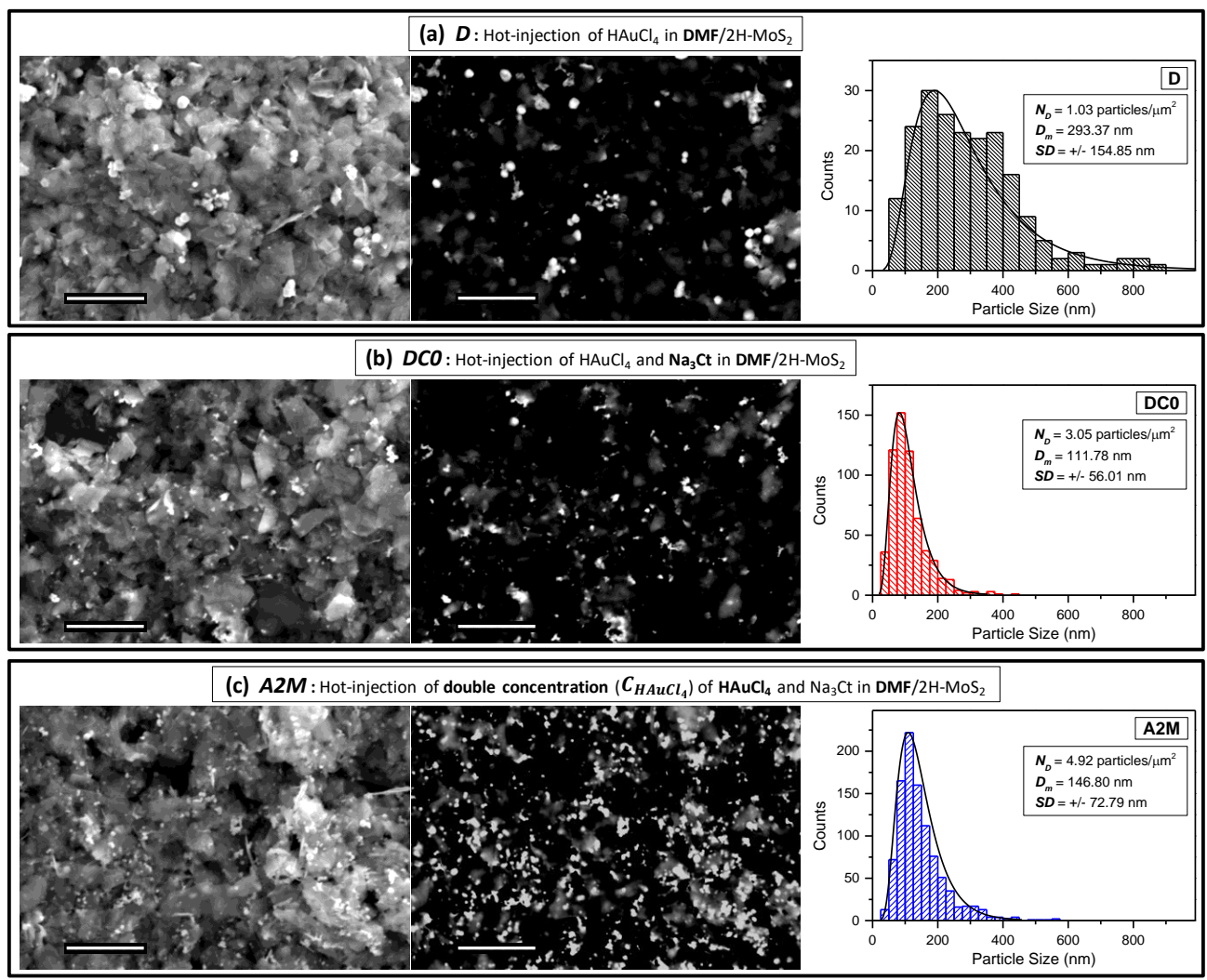

**Figure 2.** Scanning electron microscopy images (SEI, left panel), corresponding backscattered electron images (BEI, center panel), and statistical analysis of morphological parameters (right panel). AuNPs@2H-MoS$_2$ hybrids were produced in DMF by hot-injection of (a) gold precursor (HAuCl$_4$) (*D*); (b) gold precursor and citrate (*DC0*); and (c) double concentration of gold precursor ($C_{HAuCl_4}$) and citrate (*A2M*). The right panel represents particle size histograms of AuNPs for the AuNPs@2H-MoS$_2$ hybrids. $N_D$: AuNPs number density, $D_m$: mean diameter of AuNPs, and **SD**: standard deviation of NP-size. Statistical analysis has been performed on at least 3 or 4 independent SEM and corresponding BEI images for each sample. (Image Scale bar, 2 μm).

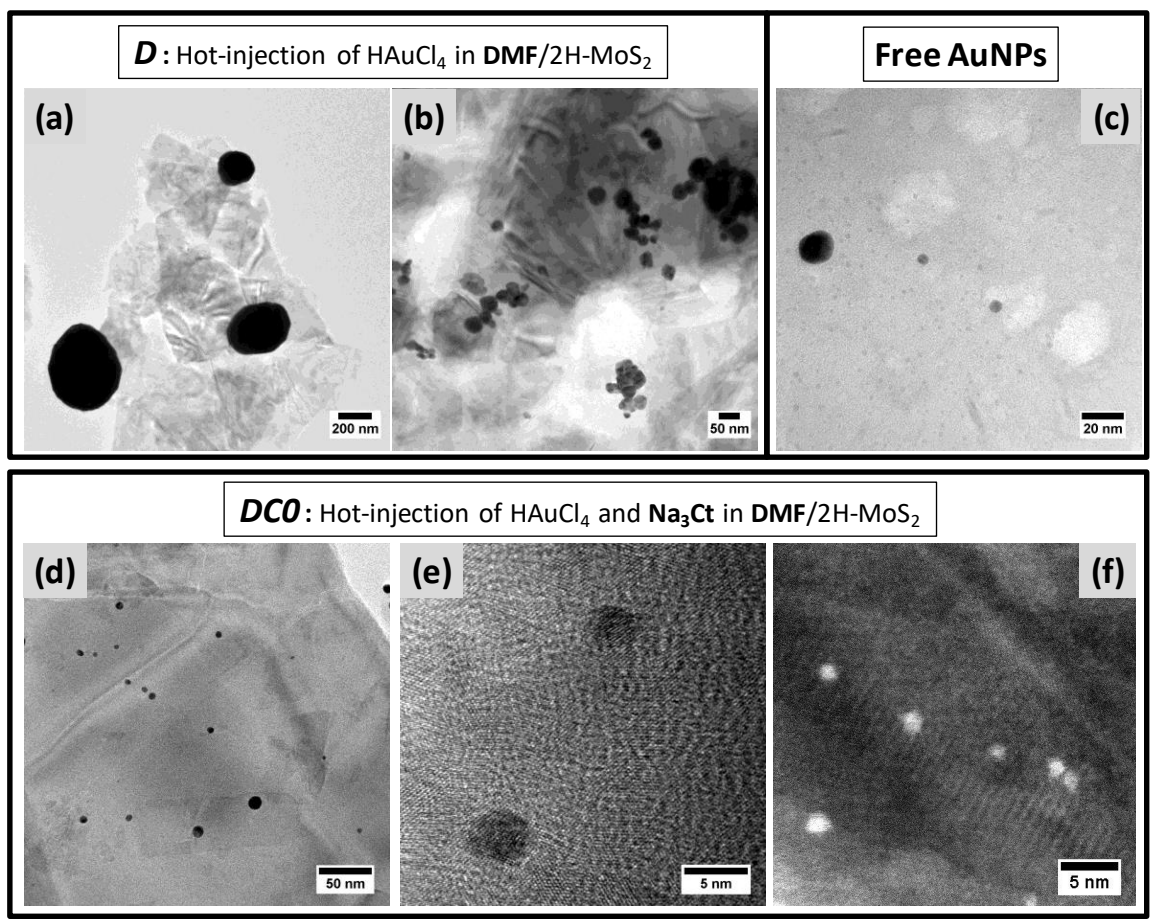

**Figure 3.** Transmission electron microscopy (TEM) images: (a-b) AuNPs@2H-MoS$_2$ hybrids synthesized in DMF by hot-injection of gold precursor (*D*); (c) "free" AuNPs formed without anchoring to 2H-MoS$_2$ NSs during the synthesis of D; and (d-e) AuNPs@2H-MoS$_2$ hybrids synthesized in DMF by hot-injection of gold precursor and citrate (*DC0*). (f) Scanning transmission electron microscopy (STEM) image demonstrating the nucleation of finer AuNPs on 2H-MoS$_2$ NSs.

As reported earlier, DMF can reduce HAuCl$_4$ to generate Au nanocrystals ($Au^0$), according to the following reaction:[21, 26-27]

Dissolution of HAuCl$_4$ in water:[23]

$$HAuCl_4 + H_2O \leftrightarrow H_3O^+ + AuCl_4^- \tag{1}$$

Hydrolysis of $AuCl_4^-$ by DMF:[21, 26-27]

$$3HCON(CH_3)_2 + 2AuCl_4^- + 3H_2O \rightarrow 2Au^0 + 3(CH_3)_2NCOOH + 6H^+ + 8Cl^- \qquad (2)$$

Interestingly, the presence of water is found to be critical.[26-27] Choi et al[27] have shown that water can accelerate the reduction reaction rate to $Au^0$. However, in our study, the water required to speed the reaction was limited, since it was supplied only through the injection of aqueous HAuCl$_4$ solution.

It should be emphasized that the above-mentioned reaction mechanism has been mainly proposed for the synthesis of AuNPs in a free metallic form in DMF solution.[19-20] This has been confirmed by the formation of free AuNPs, with an average diameter of 5 nm, not adherent to 2H-MoS$_2$ NSs, displayed in Fig. 3c. The formation of free AuNPs indicates that the nucleation and growth of AuNPs, in pure DMF, typically follow a template-free reaction route, independent of 2H-MoS$_2$ NS supports. Hence, as expected, the reported redox chemistry[6, 12-14] of $Au^+$ ions with semiconducting 2H-MoS$_2$ NSs becomes highly ineffective for our defect-free on basal plane mechanically exfoliated 2H-MoS$_2$ NSs.

For *D* synthesis, injection of gold precursor in DMF medium was executed at 140 ºC. However, dissolution of HAuCl$_4$ and associated formation of $AuCl_4^-$ ions start at much lower temperatures of ~100 ºC, as previously reported for aqueous media.[23] It can be expected that at the elevated temperature of 140 ºC, the AuNP growth rate becomes much faster than nucleation of $Au^0$ on 2H-MoS$_2$ NSs; as a result the rapid growth and aggregation of NPs is preferred over the creation of new nucleation-sites and hence gold NPs aggregate and/or form rapidly in enlarged sizes. This assumption is in agreement with the observed aggregation and formation of large NPs independent of 2H-MoS$_2$ NSs supports. The lower *N$_D$* of AuNPs observed for *D* hybrids further supports the claim (Figs. 2a and 3(a-b), *D*).

***Effect of Solvent.*** The choice of solvent is a critical factor for the chemical synthesis of AuNPs@MoS$_2$ hybrids. As discussed earlier, the most established protocols of AuNPs@MoS$_2$ hybrid synthesis are water-based utilizing the redox chemistry between metallic M-MoS$_2$ and Au-precursor.[2, 5] From our study, we found that the 2H-MoS$_2$ NSs suffer from serious aggregation, when water is used as a solvent for the synthesis of AuNPs@2H-MoS$_2$ hybrids (*W*, Fig. S1a). In contrast, using DMF as a solvent, keeping all other synthesis-conditions the same, the AuNPs@2H-MoS$_2$ hybrids (*D*, Fig. S1b) exhibits well-dispersed nature similar to the pristine 2H-MoS$_2$ NSs.

**Effect of Citrate, Na$_3$Ct, as a Secondary Reducing and Stabilizing Agent.** In the second step (Step#2), of our synthesis route (*DC0*, Fig. 1 and Table S1), a freshly prepared aqueous solution of Na$_3$Ct was injected, immediately after the addition of HAuCl$_4$ solution in the hot DMF (Step#1). The final concentration of Na$_3$Ct was maintained at $\boldsymbol{C_{Na_3Ct}}$ = 1 vol%, following the well-established Na$_3$Ct reduction protocol of gold colloidal solution.[22-23]

Citrate-reduced AuNP synthesis by the Turkevich method[22] can be described as follows:[23]

$$3C_6H_5O_7^{3-} + 2AuCl_4^- \rightarrow 2Au^0 + 3C_5H_4O_5^{2-} + 3H^+ + 8Cl^- \quad (3)$$

Figure 2b reveals the benefits of Na$_3$Ct-injection, as a secondary reducing and stabilizing agent, for the synthesis of AuNPs@2H-MoS$_2$ hybrids (*DC0*). The synthesized AuNPs become finer with smaller size ($\boldsymbol{D_m}$ ≈ 111 nm, with notably reduced $\boldsymbol{SD}$ > ±56 nm) and larger packing density ($\boldsymbol{N_D}$ ≈ 3.05 particles/μm$^2$). These finding are corroborated by TEM observations, shown in Figures 3(d-e), which reveal that the addition of Na$_3$Ct, swiftly after the injection of HAuCl$_4$ solution (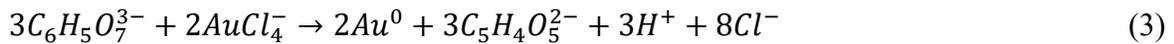$\boldsymbol{t_{Na_3Ct}}$ ≈ 1-2 min), markedly reduced the NP-size (average diameter of 5 nm) and increased number density of AuNPs. Further evidence is provided by STEM characterization, where ultra-small AuNPs with an average diameter of ≈1 nm can be clearly observed on the 2H-MoS$_2$ NSs (*DC0*, Fig. 3f).

Noticeably, a literature survey reveals that the well-established Turkevich Na$_3$Ct reduction protocol[22-23] of gold colloidal solution has been practiced mainly in aqueous medium.[22-23] So far, only one report by Choi et al,[27] has adopted the seed-mediated growth method in a mixture of DMF-water medium at 70 °C using PVP and Na$_3$Ct as a stabilizer to obtain Au rhombic dodecahedra with uniform and controllable morphology. It was declared that Na$_3$Ct served as stabilizer for the (110) facets of Au nanocrystals by adsorbing on their surfaces,[27] preventing their aggregation.

In our case of hot-injection DMF-based synthesis route at 140 °C, we believe that the Na$_3$Ct played the dual role of both *reducing* and *stabilizing* agent. When Na$_3$Ct is absent from the reaction mixture, the *D* hybrid exhibits relatively non-uniform size-distribution with formation of large AuNPs due to the aggregation (Figs. 2a and 3a-b) and "free" metal-particle formation of AuNPs (Fig. 3c). However, in the presence of Na$_3$Ct, the AuNP-distribution on the 2H-MoS$_2$ (*DC0*) reveals an obvious increase in $N_D$ (≈ 3.05 particles/μm$^2$), illustrating the enhanced rate of HAuCl$_4$-reduction, hence reflecting the reducing role Na$_3$Ct. Simultaneously, the achievement of finer NPs on *DC0* ($D_m$ ≈ 111 ± 56 nm), compared to *D* ($D_m$ ≈ 293 ± 154 nm, with $N_D$ ≈ 1.03 particles/μm$^2$), demonstrates the stabilizing function of Na$_3$Ct, which is further supported by the well-reported favorable (and selective) affinity of citrate ions towards $Au^0$.[27-28]

It has been found that use of excess citrate could lead to an uncontrolled reducing activity of Na$_3$Ct, as observed for *D2C* hybrids (synthesized by doubling the amount of Na$_3$Ct, $C_{Na_3Ct}$ = 2 vol%). Evidently, a sudden change in morphology was observed for *D2C* (Fig. S2) revealing a self-assembled nanoflake formation (Fig. S2, additional discussion in the *SI*). Since the main focus of this report was to establish a well-controlled route for the synthesis of AuNPs@2H-MoS$_2$ hybrids, while maintaining the crystalline and chemical structure of the defect-free semiconducting 2H-MoS$_2$ NSs, no further emphasis was given in this novel structure.

**Effect of HAuCl$_4$-injection time, $t_{HAuCl_4}$.** To further understand the formation mechanism of the hybrids, the sequence by which the constituents were inserted and heated in the DMF solution was investigated. As described previously, for *DC0*, both HAuCl$_4$ and Na$_3$Ct were injected just after the MoS$_2$-DMF solution reached the desired temperature of 140 ºC ($t_{HAuCl_4}$ = 140 ºC). On contrary for *DC1* hybrid, the gold-precursor was mixed with MoS$_2$-DMF solution at room temperature, prior to the heating, whereas the Na$_3$Ct was introduced only after the temperature of HAuCl$_4$-MoS$_2$-DMF mixture reached the temperature of 140 ºC. Comparative SEM study (Fig. S3) reveals that the *DC1* leads to the formation of larger AuNPs, compared to *DC0* (Fig. 2b), indicating an improvement in the Au/MoS$_2$-synthesis process by injecting HAuCl$_4$ in hot DMF solution.[19] Furthermore, these findings support further our suggestion that Na$_3$Ct acts as a secondary reducing and stabilizing agent.

For *DC1*, the reduction of HAuCl$_4$ to $Au^0$ started much earlier, as soon as the temperature crossed ~100 ºC,[23] as discussed earlier. Also, thermal decomposition of DMF into dimethylamine and carbon monoxide,[29] through the formation of unstable carbonic acid, can also start ~100 ºC and Kawasaki et al[19] suggested that this thermally generated carbon monoxide might accelerate the reduction of $AuCl_4^-$ ions. Taking into account the above arguments, it can be expected that the AuNPs have already started to form and grow in an enlarged size, before the admission of Na$_3$Ct, which results to non-uniform distribution of NPs and uncontrolled aggregation, similarly to *D* (Fig. 2a).

**Effect of HAuCl₄ concentration, $C_{HAuCl_4}$.** In order to study the effect of gold precursor concentration the HAuCl$_4$:MoS$_2$ ratio was increased by doubling the $C_{HAuCl_4}$ to 2 mM, for *A2M* hybrid, keeping all other parameters the same as those for *DC0*. It is evident that *A2M* (Fig. 2c) exhibits much higher number density of AuNPs ($N_D \approx 4.92$ particles/µm$^2$), compared to both *DC0* (Fig. 2b) and *D* (Fig. 2a) hybrids. The AuNP-growth-rate increases with the amount of Au-precursor ($C_{HAuCl_4}$) resulting to enlargement of NP-size ($D_m \approx 146 \pm 72$ nm) (Fig. 2c).

This initial study opens up the possibility for further optimisation of the size-distribution and number density of AuNPs synthesised on 2H-MoS$_2$ NSs. We believe that such optimisation is possible by fine tuning the synthesis parameters (e.g. reaction time ($t_{Rxn}$), the relative concentrations of reactants ($C_{MoS_2}$, $C_{HAuCl_4}$ and $C_{Na_3Ct}$), and possibly the reaction temperature. Nevertheless, it is worth mentioning that the current study aims at establishing an organic solvent based synthesis route for achieving the desirable hybridization of metal NPs on the basal plane of 2H-TMDC NSs; whereas at the same time maintains the semiconducting crystalline quality of TMDC NSs and attains a good dispersion of the hybrids (NPs@2H-TMDC) in the organic solvent. In this regard, the remaining of the report focuses on basic characterization studies to ensure the above-mentioned claims, rather than performing extended optimisation studies on the AuNPs@2H-MoS$_2$ hybrid synthesis.

**Chemical Stability of 2H-MoS$_2$ NSs and AuNPs in Nano-Hybrids.** Physical characterization of the nanohybrids revealed that 2H-MoS$_2$ NSs can retain their semiconducting crystalline structure, which indicates that the hybridization of 2H-MoS$_2$ NSs with AuNPs does not affect the structural properties of semiconducting 2H-MoS$_2$ NSs. In addition, the as-synthesized AuNPs are chemically stable in hot-injection DMF-Na$_3$Ct based HAuCl$_4$-reduction reaction.

***Optical Characterisation.*** The optical properties of pristine 2H-MoS$_2$ NSs and AuNPs@2H-MoS$_2$ hybrids, well-dispersed in DMF solution (≈1 mg/ml), were investigated by UV−visible absorption spectroscopy (Figs. 4 and S4). As observed from Figure 4, both pristine NSs and hybrids exhibit characteristic fingerprints of well exfoliated MoS$_2$ of 2H type with trigonal prismatic coordination.[17] Two well-defined long-wavelength peaks centered at 632 and 691 nm, attributed to the direct excitonic transitions at the K point of the Brillouin zone, do not show any obvious effect of AuNP-hybridization. However, the broader high-energy peaks around 433 and 512 nm, assigned to the direct transition from the deep valence band to the conduction band, are found to be rather affected after the hybrid synthesis. However, in a broad perception, it can certainly be concluded that the absorption spectrum of *D* hybrid is governed by the semiconducting 2H-MoS$_2$, exhibiting characteristics very similar to that of the pristine exfoliated 2H-MoS$_2$ NSs (*M*), indicating a reliable chemical stability of the product from the DMF-based Au-synthesis reaction.

Interestingly, with the introduction of Na$_3$Ct (*DC0*), significant changes can be observed in Figure 4. The peak-maximum of the broader high-energy band (in the range of 450~600 nm) is red-shifted around 545 nm, most probably due to the appearance of the characteristic surface plasmon resonance band of Au metal (≈ 565 nm), as evidenced from the spectrum of pure AuNPs in DMF solution (abbreviated as *A*, in Fig. 4). As expected, with the increase of Au-content on 2H-MoS$_2$ NSs, for *A2M* , the contribution of Au-plasmonic signal around 550 nm is enhanced (Fig. 4), revealing the existence of strong electronic interaction between AuNPs and 2H-MoS$_2$ NSs. The claim is further supported from the absorption spectrum of *D2C*, observed in Figure S4, exhibiting vividly stronger appearance of Au plasmonic peak.

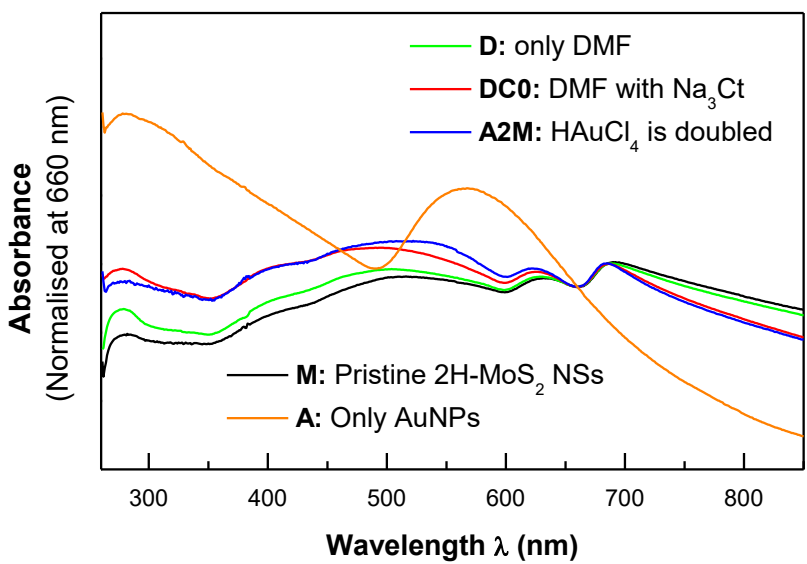

**Figure 4.** UV-visible absorption spectra of pristine semiconducting 2H-MoS$_2$ NSs (M), AuNPs (*A*), and AuNPs@2H-MoS$_2$ hybrids produced in DMF by hot-injection of (i) gold precursor (*D*); (ii) gold precursor and citrate (*DC0*); and (iii) double concentration of gold precursor and citrate (*A2M*). All samples were well-dispersed in DMF solution (≈ 1 mg/ml); pure DMF is used as a reference solution for background correction. All spectra are normalized at the wavelength of 660 nm.

With the introduction of Na$_3$Ct (*DC0*), an obvious protuberance in absorbance can be observed in lower-wavelength-regime (around 300 – 400 nm). The broad absorption at this energy-range is assigned to the direct transition from the deep valence band to the conduction band of MoS$_2$. Notably, for *A2M*, in spite of its obviously higher Au-content (observed in Fig. 2c) than that of *DC0* (Fig. 2b), its absorption spectrum remains significantly unaffected within the wavelength-range of 300 – 400 nm, when compared with that of *DC0* (Fig. 4).

*Structural Characterisation.* Structural characterizations were performed by powder X-ray diffraction (XRD) measurements (Figs. 5 and S5) on samples prepared by drop-drying DMF-dispersed solution of pristine and hybrids, on clean Si substrates. As observed from Figure 5b, the XRD patterns of *A2M* hybrid can confirm the co-existence of both exfoliated 2H-MoS$_2$ NSs and AuNPs, when they are compared with the pristine exfoliated semiconducting 2H-MoS$_2$ NSs (*M*, Fig. 5a) and pure AuNPs synthesized in DMF solution (*A*, in Fig. S5c) spectra.

Notably, in all hybrids (Figs. 5b and S5a-b), the 2H-MoS$_2$ nanosheets retain their crystalline structure with a distinctive intense (002) peak corresponding to an interlayer d-spacing of 0.614 nm, along with several weak characteristic peaks from the (100), (101), (102), (103), (006), (105), and (008) planes of polycrystalline semiconducting 2H-MoS$_2$.[17] This observation supports the non-destructive nature of our DMF-Na$_3$Ct based HAuCl$_4$-reduction reaction approach. The reduction of the Au-precursors to AuNPs and their distribution over the 2H-MoS$_2$ NSs is evidenced from four strong diffraction peaks (Figs. 5b and S5) associated with the (111), (200), (220), and (311) planes of Au (JCPDS 04-0784).[2, 30] Importantly, the domination of Au(111) peak is quite obvious for *A2M,* however the same peak is reduced for the *DC0* hybrid, in agreement with literature reports, which suggest that the citrate ion has selective affinity for Au(111).[27-28]

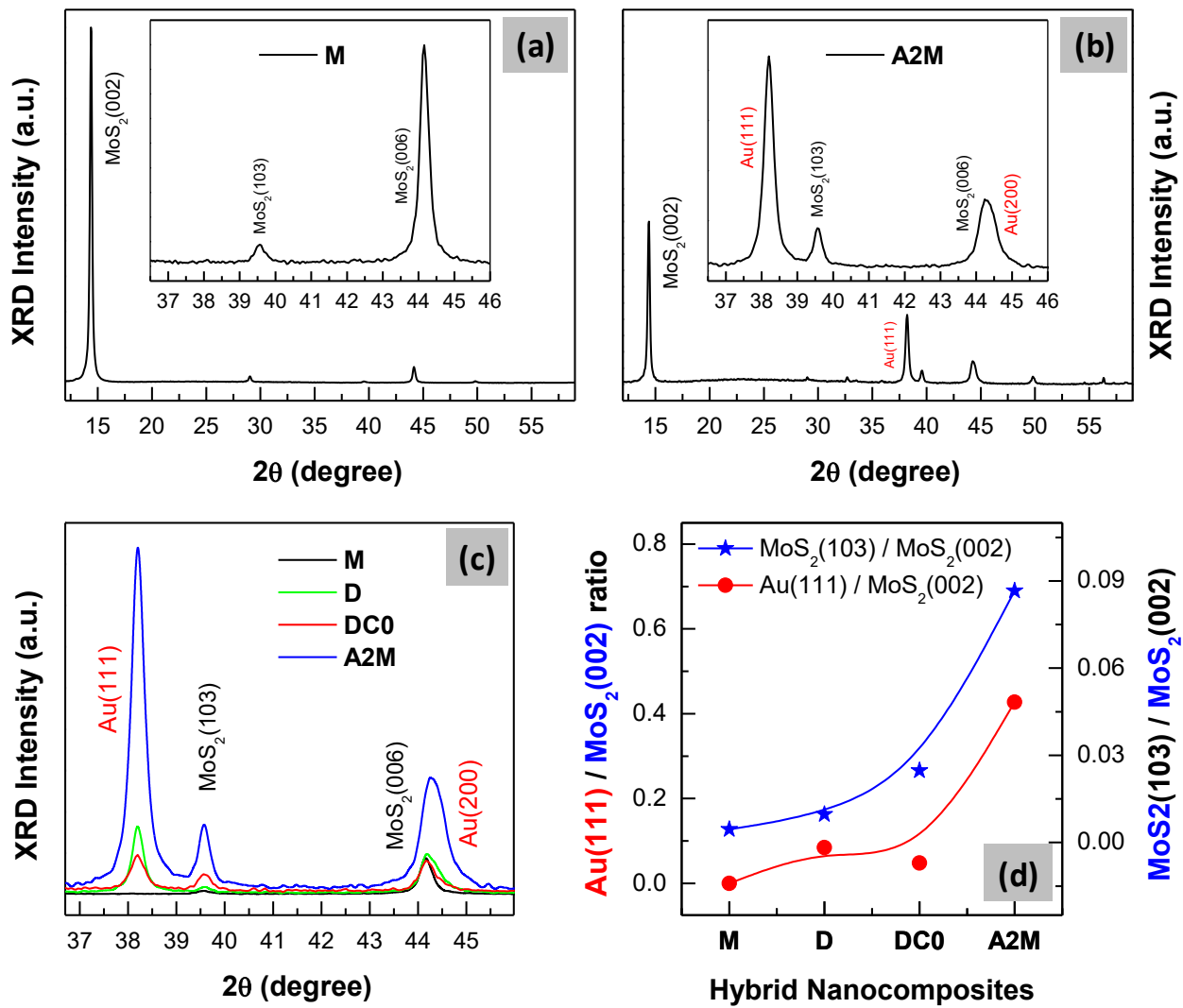

**Figure 5.** XRD spectra of (a) pristine semiconducting 2H-MoS$_2$ NSs (*M*) and (b) AuNPs@2H-MoS$_2$ hybrids synthesized in DMF by hot-injection of double concentration of gold precursor and citrate (*A2M*); respective insets illustrate the enlarged sections of XRD peaks labeled with the matched crystalline planes of semiconducting 2H-MoS$_2$ (black font) and Au (red font). (c) Comparison of XRD spectra from AuNPs@2H-MoS$_2$ hybrids produced in DMF by hot-injection of (i) gold precursor (*D*), and ii) gold precursor and citrate (*DC0*); along with *M* and *A2M*. All spectra are normalized at MoS$_2$(002) peak. (d) Comparison of the relative intensity of Au(111) and MoS$_2$(103) peaks, with respect to the MoS$_2$(002) peak for the four samples.

***Characterisation of Chemical Composition.*** X-ray photoelectron spectroscopy (XPS) was used to determine the chemical composition and bonding configurations on the same samples prepared on Si substrate. Success of our strategy for hot-injection DMF-based synthesis of AuNPs on 2H-MoS$_2$ nanosheets is clearly evidenced from the XPS analysis (Figs. 6 and S6–S7), exhibiting Mo, S and Au as the main elemental compositions. However, the presence of C and O elements cannot be ignored, which mainly originates from the solvent and the atmosphere; even for the pristine 2H-MoS$_2$ NSs (*M*), as reported in our earlier study.[17] Elemental quantification was performed after the Shirley background correction of all high resolution XPS spectra and calibration of the binding energies with reference to the C 1s line at 284.5 ± 0.2 eV associated with graphitic carbon. Calculated from the integrated areas of respective high resolution XPS spectra, the stoichiometric ratio of Mo to S is found to close to 1:2, demonstrating the expected semiconducting 2H phase of MoS$_2$, for all the AuNPs@2H-MoS$_2$ hybrids (Figs. 6 and S6).

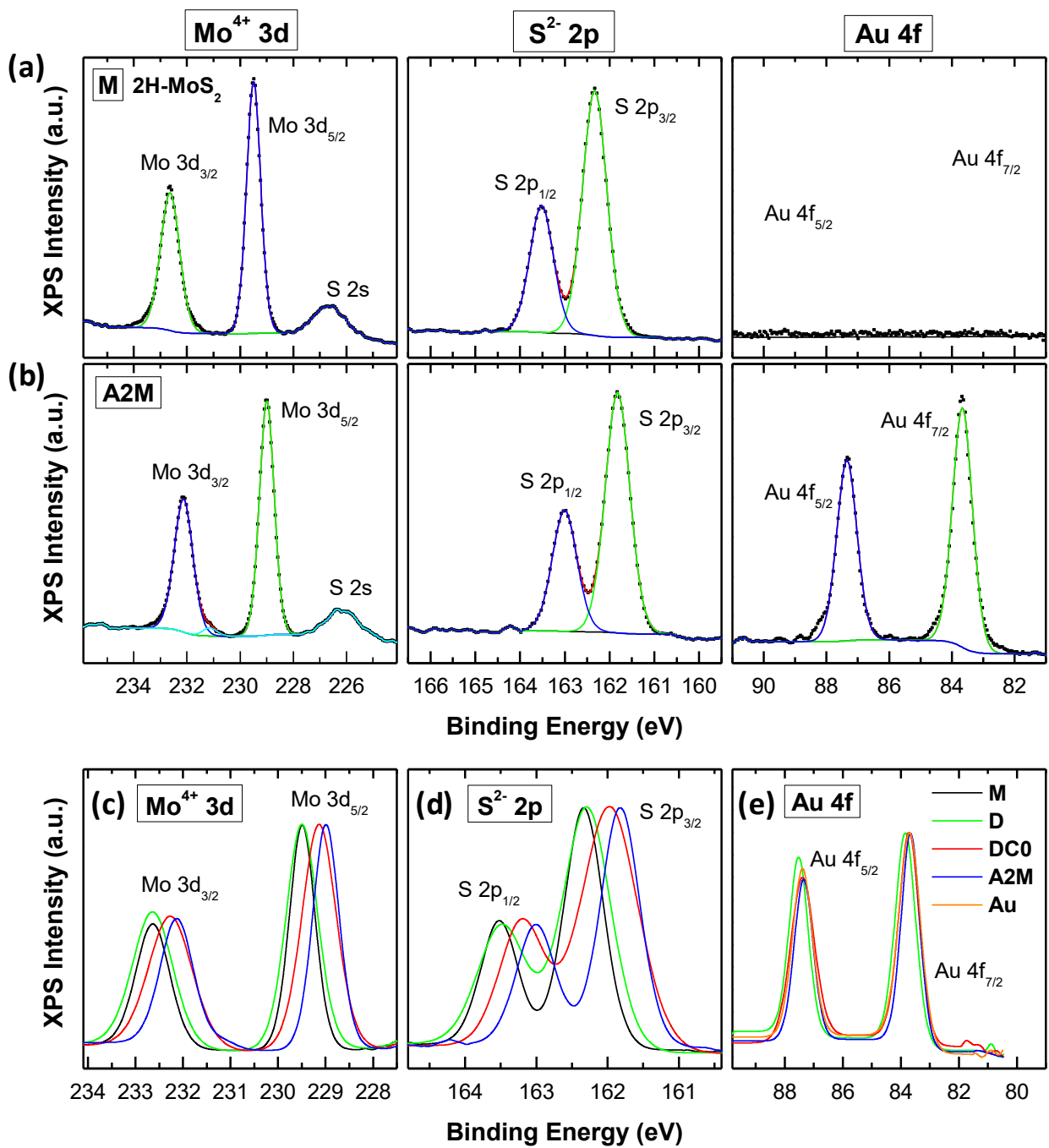

**Figure 6.** High-resolution Mo 3d (left), S 2p (middle), and Au 4f (right) XPS spectra of (a) pristine semiconducting 2H-MoS$_2$ NSs (*M*) and (b) AuNPs@2H-MoS$_2$ hybrids synthesized in DMF by hot-injection of double concentration of gold precursor and citrate (*A2M*). (c-d) Obvious red-shift of (c) Mo 3d and (d) S 2p XPS peaks reveal p-type doping of 2H-MoS$_2$ NSs (*M*) by AuNP hybridization; hybrids were synthesized in DMF by hot-injection of (i) gold precursor (*D*); (ii) gold precursor and citrate (*DC0*);

and (iii) double concentration of gold precursor and citrate (*A2M*). (e) Comparison of the respective high-resolution Au 4f XPS spectra is displayed, using that of pure AuNPs (*A*) synthesized in DMF solution as a reference. All spectra are corrected by Shirley background and calibrated with reference to the C 1s line at 284.5 ± 0.2 eV associated with graphitic carbon.

Comparison of all AuNPs@2H-MoS$_2$ XPS spectra (Fig. 6b, also Fig. S6) is found to be governed by doublet peaks of Mo 3d and S 2p XPS signals, which are similar to pristine 2H-MoS$_2$ NSs (Fig. 6a).[5, 17] The Mo 3d XPS spectrum (Fig. 6c) shows doublet peaks around 229 and 232 eV attributed to Mo$^{4+}$ 3d$_{5/2}$ and Mo$^{4+}$ 3d$_{3/2}$ orbitals, respectively. Doublet peaks around 162 and 163 eV, observed in Figure 6d, belong to S$^{2-}$ 2p$_{3/2}$ and S$^{2-}$ 2p$_{1/2}$ orbitals, respectively. These peak positions are indicative of Mo$^{4+}$ and S$^{2-}$ oxidation states in 2H phase of MoS$_2$, similar to the pristine exfoliated 2H-MoS$_2$ NSs. The results indicate the chemical stability of our 2H-MoS$_2$ NSs in the DMF-Na$_3$Ct based Au-synthesis reaction, supporting the optical (Fig. 4) and XRD results (Fig. 5) mentioned earlier. Figure 6b (also, Fig. S6) shows the Au 4f spectrum, and the doublet peaks around 87.3 (Au 4f$_{5/2}$) and 83.7 eV (Au 4f$_{7/2}$) provide direct evidence for the reduction of the Au-precursors hence the formation of AuNPs on 2H-MoS$_2$ NSs.[5, 31]

Here, it is worth to mention, that the uncontrolled reducing activity of citrate, when used in excessive amounts, is vividly evidenced from the XPS analysis *D2C* hybrid (synthesized by doubling the amount of Na$_3$Ct, Figs. S6d and S7c). All three Mo 3d, S 2p and Au 4f XPS signals of *D2C* (Fig. S6d) appear with distorted shape due to excessive contamination. The carbon-contamination for *D2C* also becomes significant and complex in nature (Fig. S7c), when compared to pristine 2H-MoS$_2$ NSs (*M*) (Fig. S7a) and other hybrids (Fig. S7b).

**Evidence of Hybridization of AuNPs with 2H-MoS₂ NSs.** Detailed analysis of XRD and XPS results provided evidence not only on the successful of AuNPs synthesis on mechanically exfoliated 2H-MoS₂ NSs, but also on their highly desired Au-MoS₂ hybridization. Both XRD and XPS studies clearly reveal that such hybridization can be observed more extensively for the AuNPs@2H-MoS₂ hybrids synthesized in presence of Na₃Ct (*DC0*), compared to those synthesized in pure DMF i.e. without Na₃Ct (*D*).

*Preferential Crystal Plane Growth of AuNPs on 2H-MoS₂ NSs.* XRD results, presented in Figures 5b and S5, revealed that the crystalline structure of the synthesized AuNPs decorated on the exfoliated 2H-MoS₂ NSs is mainly dominated by the Au(111) face, due to the selective affinity of the citrate ions for Au(111).[27-28] As a result, low-index Au(111) facets are enlarged at the expense of high-index facets like Au(100) and Au(110) during crystal growth. The *A2M* nanohybrids, with high $C_{HAuCl_4}$, exhibit significant increase in Au(111) peak, relative to the second-intense peak from Au(200) planes (Fig. 5c).

Importantly, as previously reported, the Au(111) has the lowest lattice mismatch with MoS₂(001), which would also promote the preferential orientation of Au(111) crystal planes during AuNPs-growth on semiconducting 2H-MoS₂ surface.[8] It can be evidenced from Figure 5d, which presents the peak intensity ratio of Au(111) relative to MoS₂(002) peak, that the Au(111) peak intensity is markedly enhanced for the (*A2M*) hybrids with increased gold concentration. Interestingly, the MoS₂(103) peak, the second-intense peak of 2H-MoS₂ NSs, reveals an unique structural evolution depending on Au-synthesis conditions. As observed from Figure 5d, with the increase in Au-content, the ratio of MoS₂(103) peak, relative to MoS₂(002) peak, is enhanced in similar fashion as the ratio of Au(111) peak relative to MoS$_2$(002) peak, however much more intensely. Considering that the MoS$_2$(002) peak represents the basal plane, Figure 5d indicates the preferential affinity of AuNPs nucleation and growth on the basal plane of 2H-MoS$_2$ NSs.

**Au-doping on 2H-MoS$_2$ NSs: p-type Doping.** Interestingly, Figures 6c and 6d clearly exhibit an obvious red-shift of both Mo$^{4+}$ and S$^{2-}$ peaks to lower binding energies, for the AuNPs@2H-MoS$_2$ hybrids, relative to those of pristine exfoliated 2H-MoS$_2$ NSs (*M*), indicating a down-shift of the Fermi level in MoS$_2$ due to p-type doping.[1, 3-6, 8, 24-25, 30-31] Notably, the hybrid (*D*), synthesized in pure DMF, without Na$_3$Ct, does not show any obvious peak-shift or doping-effect. Such observation certainly supports our earlier SEM observations, which showed that the synthesis of AuNPs mainly happens in a free metallic form in DMF solution independent of 2H-MoS$_2$ NS supports. The nucleation and growth of AuNPs is spontaneous and faster in free state inside the DMF solution, and strongly preferred over $Au^0$ nucleation on 2H-MoS$_2$ NSs. The argument can also support the aggregation and formation of large NPs for the *D* hybrids independent of 2H-MoS$_2$ NSs supports (Figs. 2a and 3(a-b)). The optical (Fig. 4) and structural characterizations (Fig. S5a) of *D* also support the claim.

However, with the incorporation of Na$_3$Ct, for *DC0*, the p-doping of Au on 2H-MoS$_2$ NSs becomes highly evident (Figs. 6c and 6d). With the increase in $C_{HAuCl_4}$ (*A2M*), both Mo$^{4+}$ and S$^{2-}$ peaks exhibit the largest red-shift by ca. of 0.5 eV demonstrating a large doping effect (Figs. 6c and 6d, respectively). Here, the AuNPs act as a p-type dopant in semiconducting 2H-MoS$_2$ since the $AuCl_4^-$ ions in solution can strongly withdraw electrons from 2H-MoS$_2$ NSs and reduce to AuNPs.[1, 4, 8, 12, 24-25]

Interestingly, the representative AuNPs@2H-MoS$_2$ hybrids do not exhibit any obvious shift of Au 4f peak compared to that of the pure AuNPs synthesized in DMF solution (*A*, in Fig. 6e).

***Electrocatalytic Activity for Hydrogen Evolution Reaction (HER).*** The benefits of Au-MoS$_2$ hybridization can be demonstrated from the electrocatalytic behavior of AuNPs@2H-MoS$_2$ hybrids, as illustrated in Figures 7 and S8. The working electrode was fabricated by drop-drying the catalyst ink (DMF solution of hybrids) onto a polished glassy carbon electrode (GCE). As shown in Figure S8a, initial electrochemical characterization, via cyclic voltammetric (CV) studies in H$_2$SO$_4$ solution, the AuNPs@2H-MoS$_2$ hybrid (*A2M*) exhibits typical voltammetric characteristics of gold electrodes.[1, 32] Furthermore, by probing the Fe(CN)$_6^{3-/4-}$ redox activity,[1] it is found that the AuNPs@2H-MoS$_2$ hybrid (*A2M*) exhibits enhanced redox peak currents, relative to the pristine semiconducting 2H-MoS$_2$ NSs (*M*), indicating a higher electroactive surface area (~1.24 times higher) and superior electron transfer performance (Figs. S8b and S8c).

The linear sweep voltammograms of the AuNPs@2H-MoS$_2$ hybrids are presented in Figure 7a, which reveal a significant improvement in the electrocatalytic activities for H$_2$ generation. Due to increased hybridization, demonstrated by enhanced p doping, the *A2M* hybrid, shows the best catalytic performance exhibiting an onset potential (V$_{OC}$) of −0.17 V vs. RHE and an overpotential ($\eta_{10}$) of 337 mV at the cathodic current density of 10 mA/cm$^2$; the last is commonly used as a figure of merit for the HER performance of an electrocatalyst. Both potentials are considerably more positive than those of pristine semiconducting 2H-MoS$_2$ NSs (*M*), which display V$_{OC}$ = −0.25 V and $\eta_{10}$ = 442 mV. It is found that the AuNPs-hybridization can improve the V$_{OC}$ and $\eta_{10}$ values of 2H-MoS$_2$ electrocatalyst by 32% and 24%, respectively.

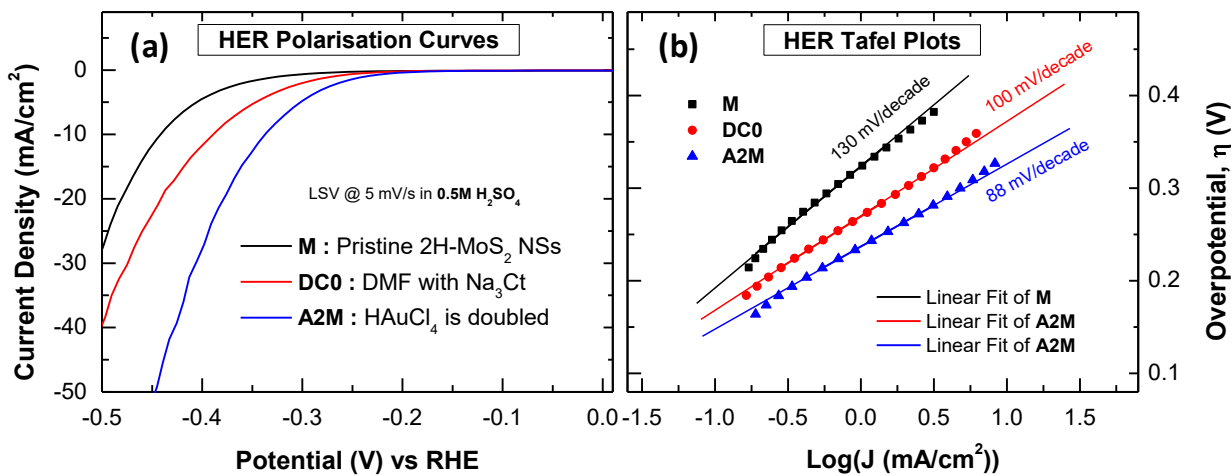

**Figure 7.** (a) Polarization curves of pristine semiconducting 2H-MoS$_2$ NSs (*M*) and AuNPs@2H-MoS$_2$ hybrids synthesized in DMF by hot-injection of (i) gold precursor and citrate (*DC0*) and (ii) double concentration of gold precursor and citrate (*A2M*), in 0.5 M H$_2$SO$_4$ at 5 mV/s vs. RHE. (b) Respective Tafel plots of overpotential (η) versus Log(J); J represents the current density.

Tafel slope is a useful metric to assess the performance of catalysts and is estimated from the linear portions of Tafel plots, satisfying the equation: η = $b$ log|J| + $a$, where η is overpotential, J is the current density, *a* is exchange current density, and *b* is the Tafel slope. The *A2M* hybrid is found to exhibit the smallest Tafel slope of 88 mV/decade, improving the pristine semiconducting 2H-MoS$_2$ (*M*) electrodes, by 32%. Tafel slope can also help to define the mechanistic reaction processes of HER.[17] Theoretically, HER proceeds through three principal reaction steps in acidic media, namely Volmer, Heyrovsky and Tafel reactions, associated with the Tafel slopes of 120, 40 and 30 mV/decade, respectively. Here, for the AuNPs@2H-MoS$_2$ *A2M* hybrid, the HER proceeds through the Volmer-Heyrovsky mechanism, with the Heyrovsky reaction as the rate-determining step, as indicated by its Tafel slope of 88 mV/decade. On the contrary, the large Tafel slope of 130 mV/decade of pristine 2H-MoS$_2$ (*M*) suggests that the HER follows mainly the Volmer reaction.

Improved electrocatalytic response to HER at the AuNPs@2H-MoS$_2$ NSs electrode, compared to pristine 2H-MoS$_2$ NSs, not only suggests its potential for the energy applications, but also ensures the success of the proposed DMF-based hot-injection synthesis route. It is expected, that further fine tuning of synthesis parameters (e.g. reaction time ($t_{Rxn}$), the relative concentrations of reactants ($C_{MoS_2}$, $C_{HAuCl_4}$ and $C_{Na_3Ct}$), also possibly the reaction temperature would improve the size-distribution and number density of AuNPs, which in turn could improve further the HER performance. The primary focus of the current report was to establish the organic solvent hot-injection synthesis route for the controlled synthesis of AuNPs@2H-MoS$_2$ hybrids; no further emphasis was given in the HER application.

- **CONCLUSIONS**

We report on the controlled synthesis of gold nanoparticles (AuNPs) on semiconducting transition-metal dichalcogenide (TMDC) layers, which is of high technological interest in many applications including photocatalysis, optical sensing, and optoelectronics. At present, the commonly used aqueous solution approaches suffer from poor dispersion of the produced hybrids leading to stacking effects and also from very limited coverage of the TMCD basal plane with AuNPs due to absence of defect sites. We have tackled these challenges, and here we present for the first time the successful demonstration of citrate modified DMF-based hot-injection chemical synthesis of AuNPs@2H-MoS$_2$ hybrids. This organic solvent-based synthesis route eliminates problems of poor dispersion of AuNPs@2H-MoS$_2$ hybrids found in aqueous solvents and confined hybridization at the edges of 2H-MoS$_2$ NSs. Importantly, the semiconducting crystalline quality of the pristine 2H-MoS$_2$ NS is maintained in the produced hybrids. Though a systematic investigation of synthesis parameters a mechanistic understanding of their role has been obtained. The study establishes the use of trisodium citrate as the secondary reducing and stabilizing agent, which enhances the nucleation and improves the hybridization of AuNPs on 2H-MoS$_2$ NSs as revealed by the induced p-type doping. These beneficial effects of Na$_3$Ct were further evidenced by the improved electrocatalytic activity for hydrogen evolution reaction. It is expected, that fine tuning of synthesis parameters would improve further the HER performance, however such optimization is not within the scope of this work. We believe that this organic solvent synthesis approach can be adopted for other hybrid systems opening the way for controlled hybridization of semiconducting layers with metal nanoparticles.

- **ASSOCIATED CONTENT**

**Supporting Information (*SI*)**

Table presenting the list of AuNPs@2H-MoS$_2$ hybrid nanocomposites reported in this study; Additional experimental details (chemical and characterisations); Effect of solvent: DMF vs. Water; Na$_3$Ct concentration, $\boldsymbol{C_{Na_3Ct}}$, and HAuCl$_4$-injection time, $\boldsymbol{t_{HAuCl_4}}$; Additional information on the UV-visible absorption spectroscopy, X-ray diffraction, X-ray photoelectron spectroscopy, and electrochemical studies.

The following files are available free of charge.

- **AUTHOR INFORMATION**

**Corresponding Author**

*P. Papakonstantinou. E-mail: p.papakonstantinou@ulster.ac.uk.

**Funding Sources**

British Council, Newton fund Institutional Links, Ref: 216182787.

Department for Education in Northern Ireland and Ulster University (VCRS studentship)

European Community (Erasmus+ founding for international mobility)

Ministry of Education, Universities and Research in Italy (20152EKS4Y)

University of Catania (Piano della Ricerca di Ateneo 2016-2018)

**Notes**

The authors declare no competing financial interest.

- **ACKNOWLEDGMENTS**

This work was supported by the British Council (Newton fund, Institutional Links, Ref: 216182787); Department for Education in Northern Ireland and Ulster University (VCRS studentship for SS); European Community (Erasmus+ founding for international mobility, for O.T); Ministry of Education, Universities and Research in Italy (MIUR under Grant PRIN 2015-20152EKS4Y project); and University of Catania (Piano della Ricerca di Ateneo 2016-2018).

## TOC Graphic

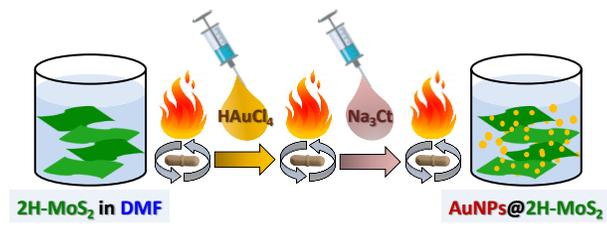

# Supporting Information

# Organic Solvent Based Synthesis of Gold Nanoparticle - Semiconducting 2H-MoS$_2$ Hybrid Nanosheets

*Abhijit Ganguly,[†] Olga Trovato,[#] Shanmughasundaram Duraisamy,[†] John Benson,[‡] Yisong Han,[†] Cristina Satriano[#] and Pagona Papakonstantinou[\*,†]*

[†]School of Engineering, Engineering Research Institute, Ulster University, Newtownabbey BT37 0QB, United Kingdom

[#] Department of Chemical Sciences, University of Catania, Viale Andrea Doria 6, 95125 Catania, Italy

[‡]2-DTech, Core Technology Facility, 46 Grafton St., Manchester M13 9NT, United Kingdom

[\*]**Corresponding Author.** E-mail address: p.papakonstantinou@ulster.ac.uk

**CONTENT**

- **List of AuNPs@2H-MoS$_2$ Hybrids**

Table S1. List of AuNPs@2H-MoS$_2$ Hybrids

| Sample Names | | MoS$_2$ in DMF/water | | | HAuCl$_4$ aq. soln. | | Final $C_{HAuCl_4}$ | Na$_3$Ct aq. soln. | | Final $C_{Na_3Ct}$ | $t_{Rxn}$ | |
|---|---|---|---|---|---|---|---|---|---|---|---|---|
| | | $M_{MoS_2}$ | Solvent | $V_{DMF}$ | $M_{HAuCl_4}$ | $t_{HAuCl_4}$ | | $M_{Na_3Ct}$ | $t_{Na_3Ct}$ | | | |
| *M* | MoS$_2$ | | | | | | | | | | | *Only Pristine 2H-**MoS$_2$ NSs*** |
| *W* | AMw | 5 mg | water | 10 ml | 3.4 mg | at 100 °C | 1 mM | | | | 30 min | *Agglomeration of* MoS$_2$ NSs |
| *D* | AMd | 5 mg | **DMF** | 10 ml | 3.4 mg | at **140 °C** (*after boiling*) | 1 mM | | | | 30 min | *Huge* **AuNPs** |
| *DC0* | AMdc0 | 5 mg | DMF | 10 ml | 3.4 mg | at 140 °C | 1 mM | 10 mg | ~2 mins | **1 vol%** | ~30 min | *Fine* **AuNPs** |
| *DC1* | AMdc-1 | 5 mg | DMF | 10 ml | 3.4 mg | at **RT** (***Step#1***) | 1 mM | 10 mg | at 140 °C (*after boiling*) | 1 vol% | ~150 min | *Large* **AuNPs** |
| *A2M* | A2Mdc0 | 5 mg | DMF | 10 ml | **6.8 mg** | at 140 °C | **2 mM** | 10 mg | ~2 mins | 1 vol% | ~30 min | *High Density Fine* **AuNPs** |
| *D2C* | AMd2c0 | 5 mg | DMF | 10 ml | 3.4 mg | at 140 °C | 1 mM | 10 mg | ~2 mins | **2 vol%** | ~30 min | *Microflowers Structures* |
| *A* | Adc0 | | DMF | 10 ml | 3.4 mg | at 140 °C | 1 mM | 10 mg | ~2 mins | 1 vol% | ~30 min | *Only* **AuNPs** |

- **Additional Experimental Details**

**Chemicals.** Molybdenum (IV) sulphide ($MoS_2$) powder (<2 μm, 99.0%), Gold(III) chloride hydrate ($HAuCl_4 \cdot xH_2O$, 99.999% metal basis), Trisodium citrate dehydrate ($Na_3Ct$, ≥99.0%), the room temperature ionic liquid (RTIL), 1-butyl-3-methylimidazolium hexafluorophosphate ($BMIMPF_6$, ≥97.0%) were purchased from Sigma-Aldrich. Solvents like N,N-dimethylformamide (DMF, ≥99.9%) and Acetone (≥99.8%) were also supplied from Sigma-Aldrich. Ultrapure water (resistivity of 18.2 MΩ·cm, Millipore) was used to prepare all aqueous solutions.

Mechanical exfoliation of semiconducting $2H\text{-}MoS_2$ platelets was performed by mechanical grinding, using a mortar grinder system (RM200, Retsch GmbH) followed by sequential centrifugation steps, using a Thermo Scientific Sorvall ST-16 Centrifuge system.

**Characterizations.** Surface morphology of all AuNPs@$2H\text{-}MoS_2$ hybrids was studied by a scanning electron microscope (SEM, FEI Quanta 200 2D) at an accelerating voltage of 15kV. Prior to basic characterizations, all as-prepared powder hybrids were well-dispersed in DMF solution (≈ 1 mg/ml), under ultrasonication. For surface morphology and elemental characterizations, well-dispersed DMF inks were drop-dried on clean Si substrates.

Transmission electron microscopy (TEM) and scanning transmission electron microscopy (STEM) images were obtained on the JEOL JEM 2011 at an accelerating voltage of 200 kV. TEM samples were prepared by drop-drying 2 μl of each nanocomposite-DMF solution onto carbon micro-grids (Agar scientific, S147-3, holey carbon film 300 mesh Cu).

Absorption spectroscopy was performed in the ultraviolet-visible (UV-vis) range (800 – 200 nm) on a Perkin Elmer Lambda 35 spectrometer. All AuNPs@2H-MoS$_2$ hybrids were well-dispersed in DMF solution ($\approx$ 1 mg/ml). Pure DMF was used as a reference solution for background correction.

X-ray photoelectron spectroscopy (XPS) was performed on samples drop-dried on cleaned Si substrates, using Kratos AXIS ultra DLD with an Al K$\alpha$ (hv = 1486.6eV) x-ray source. The elemental quantification was performed after the Shirley background correction of all high resolution spectra and calibrating the binding energy with reference to the C 1s line at 284.5 $\pm$ 0.2 eV associated with graphitic carbon.

X-ray diffraction (XRD) analysis was conducted on the same samples used for XPS studies, employing a Bruker D8-diffractomer with Cu-K$\alpha$ radiation source (40 kV, 40 mA, $\lambda$=1.5314 Å).

Electrochemical characterizations and the hydrogen evolution reaction (HER) studies were performed on an Autolab, PGSTAT/FRA system, employing a typical 3-electrodes configuration with a platinum wire (Pt, CHI. Instrument, Inc.) as counter electrode (CE) and Ag/AgCl (3M KCl, CHI. Instrument, Inc.) as reference electrode (RE). Prior to all the electrochemical characterizations and the hydrogen evolution reaction (HER) studies, the catalyst inks were prepared by dispersing well 3 mg of as-synthesized powder materials in 1 ml DMF (adding 30 μl of Nafion solution) under adequate ultrasonication. The working electrode (WE) was finally fabricated by drop-drying the catalyst ink onto a polished glassy carbon electrode (GCE, 3 mm Diameter, BASi) with a desired catalyst-loading of ~300 μg/cm$^2$.

- **Effect of Solvent: DMF vs. Water ( D vs. W)**

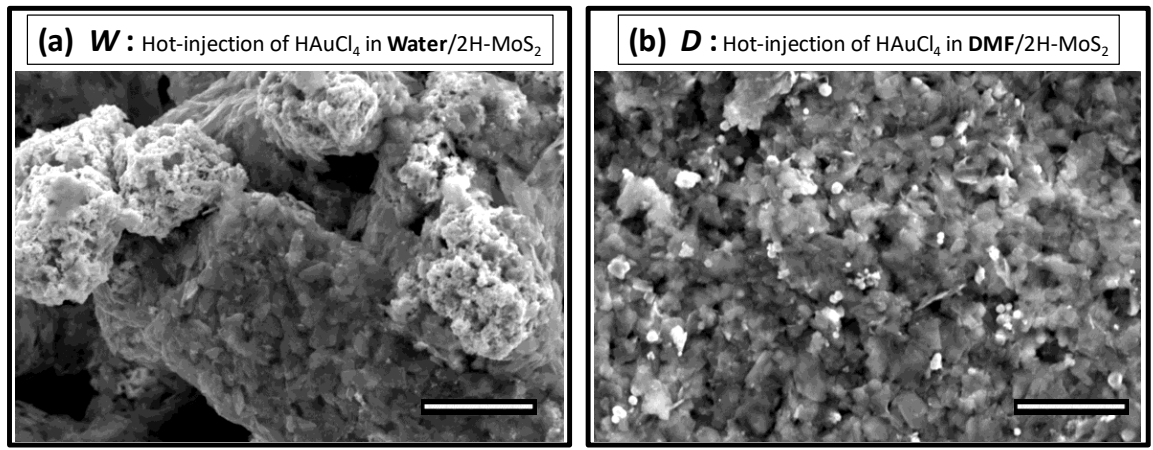

**Figure S1.** Scanning electron microscopy images (SEI) of AuNPs@2H-MoS$_2$ hybrids synthesized by hot-injection of gold precursor (a) in water (*W*) and (b) in N,N-dimethylformamide, DMF (*D*) ; all other synthesis-parameters remain the same. (Scale bar, 5 μm)

- **Effect of Na₃Ct concentration, $C_{Na_3Ct}$**

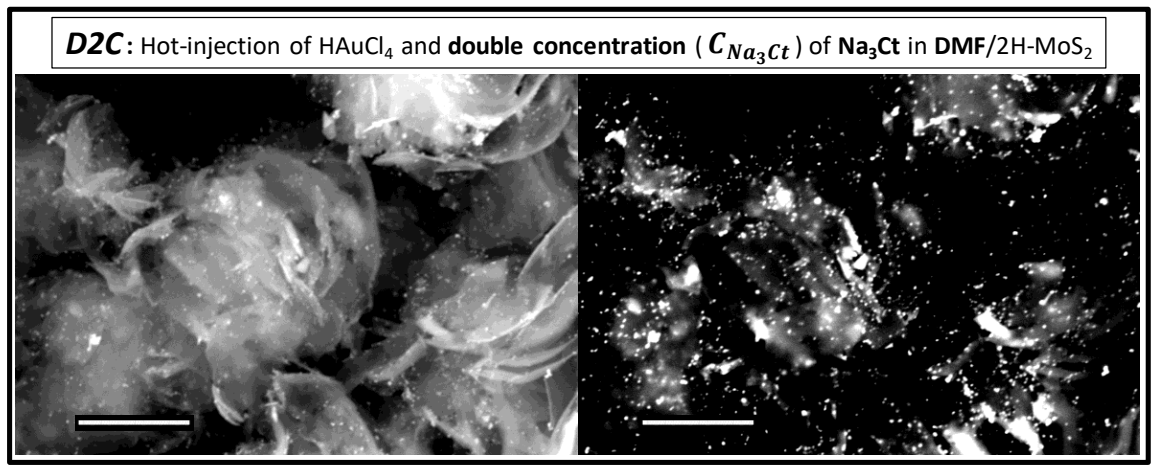

**Figure S2.** Scanning electron microscopy images (SEI) at left and corresponding backscattered electron images (BEI) at right: AuNPs@2H-MoS₂ hybrids synthesized in DMF by hot-injection of gold precursor and double concentration ($C_{Na_3Ct}$) of citrate (*D2C*). (Scale bar, 2 μm)

***Effect of Na₃Ct concentration, $C_{Na_3Ct}$:*** Na₃Ct concentration effects on the formation of hybrids were investigated by doubling the amount of Na₃Ct, ($C_{Na_3Ct}$ = 2 vol%) (*D2C*), while maintaining all other steps exactly the same as those for *DC0*. Interestingly, a sudden change in morphology was observed for *D2C* (Fig. S2a) revealing a self-assembled nanoflake formation. No report can be found on such morphological effect of DMF and citrate on MoS₂. In an earlier study, Dickmeis et al.[1] has reported the microwave-assisted decomposition of DMF into dimethylamine and carbon monoxide (maximum temperature of 230 ºC). It was suggested that, due to its nucleophilic character, the created amine leads to formation of a modified polymer.[1]

Here, we can assume that the reducing activity of Na$_3$Ct may accelerate the decomposition of DMF even at the lower temperature (140 °C).

Furthermore, chemical characterization by XPS analysis (Figs. S6d and S7c) revealed that use of excess Na$_3$Ct leads to an uncontrolled reducing activity of Na$_3$Ct, as observed from the distortion of Mo 3d, S 2p and Au 4f XPS signals of the *D2C* hybrid (Fig. S6d). The carbon-contamination also becomes significant and complex in nature for the *D2C* hybrid (Fig. S7c), when compared to pristine 2H-MoS$_2$ NSs (*M*) (Fig. S7a) and other AuNPs@2H-MoS$_2$ hybrids (Fig. S7b).

Nevertheless, further investigation on this sample (*D2C* hybrids) is considered out of scope for this work. Our focus is mainly to establish a non-destructive synthesis route for AuNPs@2H-MoS$_2$ hybrids, maintaining in pristine form the crystalline and chemical structure of our defect-free semiconducting 2H-MoS$_2$ nanosheets.

- **Effect of HAuCl$_4$-injection time, $t_{HAuCl_4}$**

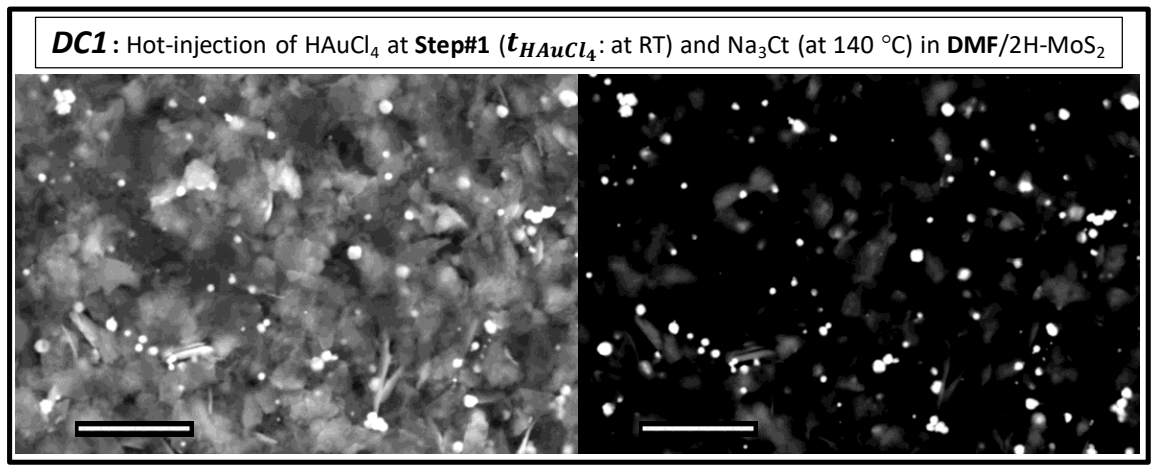

**Figure S3.** Scanning electron microscopy images (SEI) at left and corresponding backscattered electron images (BEI) at right: AuNPs@2H-MoS$_2$ hybrids synthesized in DMF by injection of gold precursor at Step#1, prior to heating ($t_{HAuCl_4}$: at RT, *DC1*). (Scale bar, 2 μm)

- **UV-visible Absorption Spectroscopy**

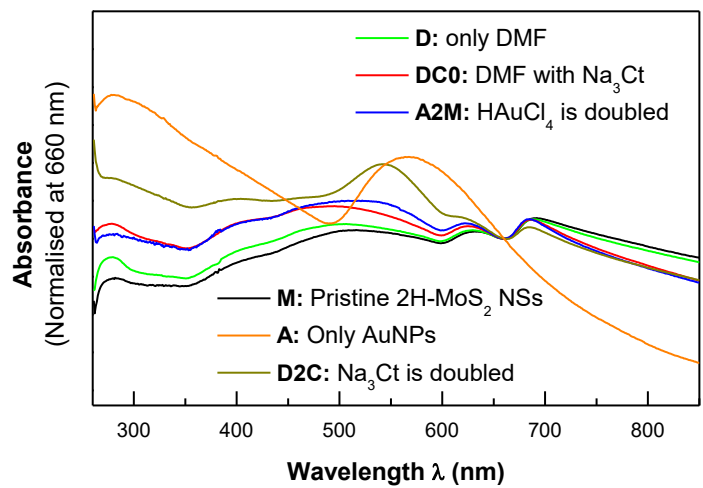

**Figure S4.** UV-visible absorption spectra of pristine semiconducting 2H-MoS$_2$ NSs (M), AuNPs (*A*), and AuNPs@2H-MoS$_2$ hybrids synthesized in DMF by hot-injection of (i) gold precursor (*D*); (ii) gold precursor and citrate (*DC0*); (iii) double concentration of gold precursor and citrate (*A2M*); and (vi) gold precursor and double concentration of citrate (*D2C*). All samples were well-dispersed in DMF solution (≈ 1 mg/ml); pure DMF is used as a reference solution for background correction. All spectra are normalized at the wavelength of 660 nm.

- **X-ray Diffraction (XRD) Study**

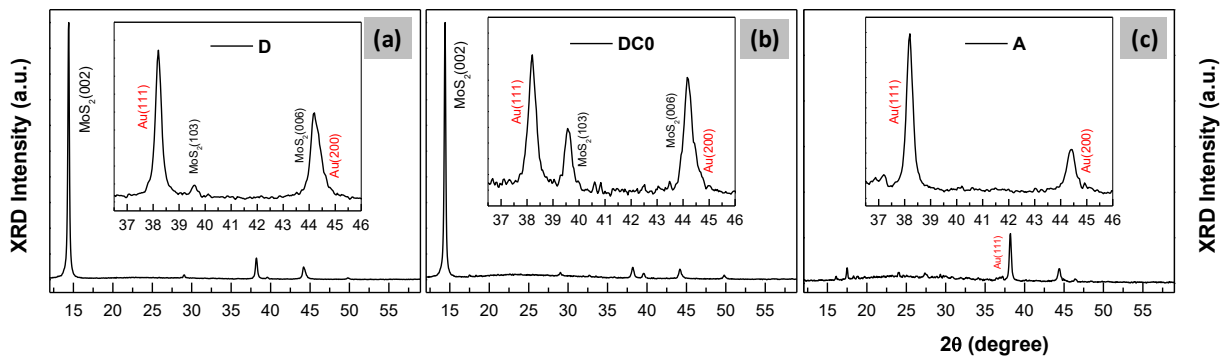

**Figure S5.** XRD patterns of AuNPs@2H-MoS$_2$ hybrids synthesized in DMF by hot-injection of (a) gold precursor (*D*); and (b) gold precursor and citrate (*DC0*). XRD pattern of AuNPs (*A*) is presented as a reference in (c). Respective insets illustrate the enlarged sections of XRD peaks labeled with the matched crystalline planes of semiconducting 2H-MoS$_2$ (black font) and Au (red font). All spectra are normalized at MoS$_2$(002) peak.

- **X-ray Photoelectron Spectroscopy (XPS)**

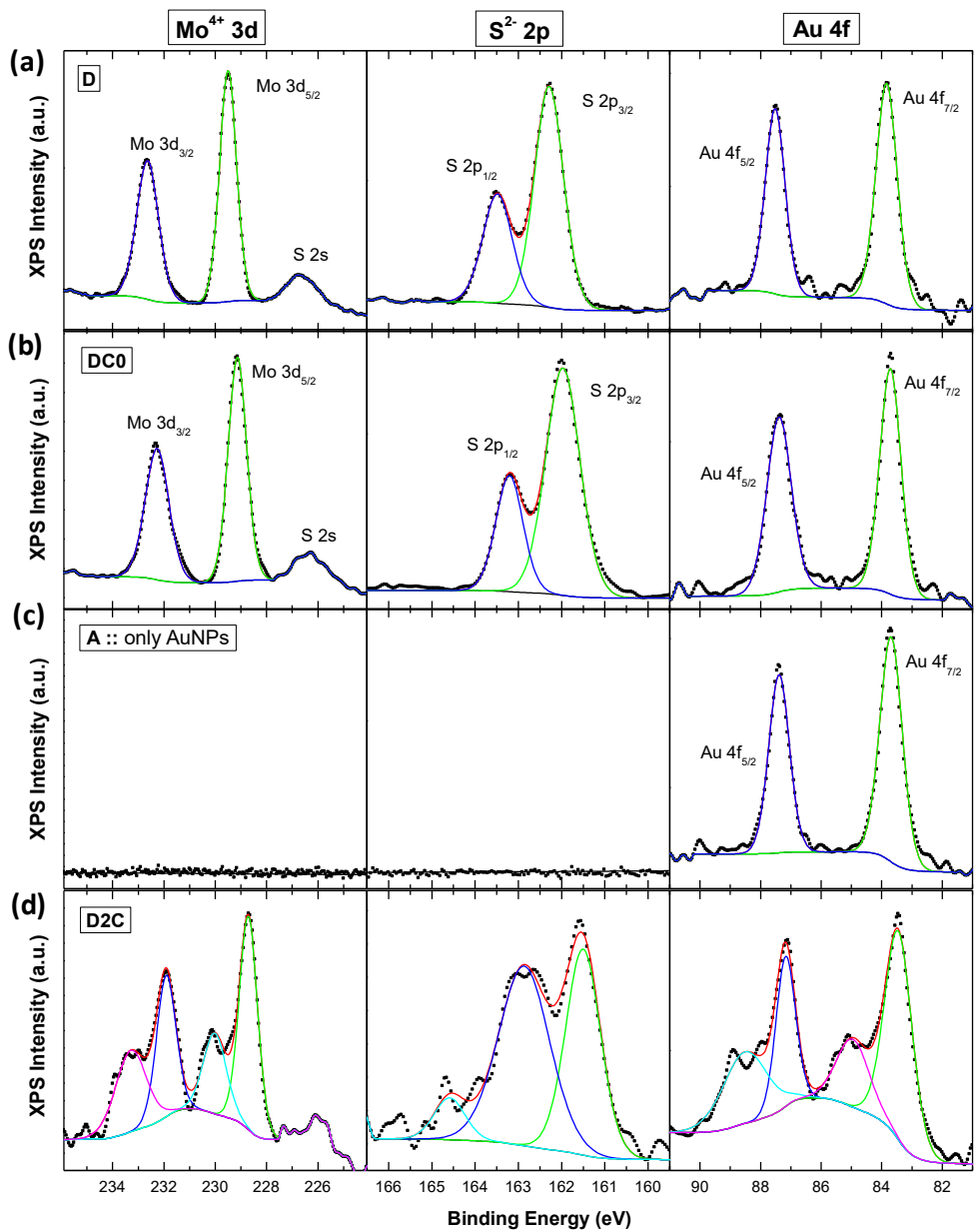

**Figure S6.** High-resolution Mo 3d (left), S 2p (middle), and Au 4f (right) XPS spectra of AuNPs@2H-MoS$_2$ hybrid synthesized in DMF by hot-injection of (a) gold precursor (*D*); (b) gold precursor and citrate (*DC0*); and (d) double concentration of gold precursor and citrate (*D2C*). Corresponding XPS spectra of AuNPs (*A*) are illustrated as a reference in (c). All spectra are corrected by Shirley background and calibrated with reference to the C 1s line at 284.5 ± 0.2 eV associated with graphitic carbon.

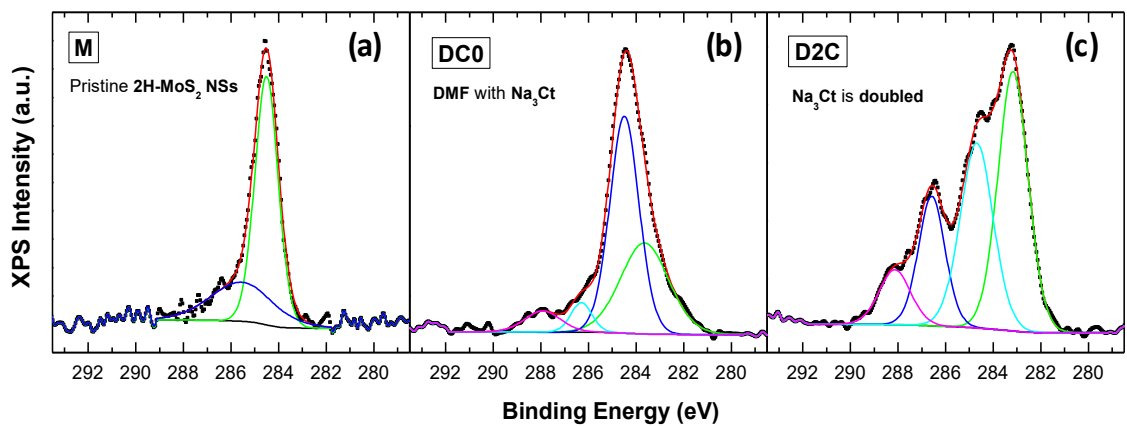

**Figure S7.** High-resolution C 1s XPS spectra of (a) pristine semiconducting 2H-MoS$_2$ NSs (*M*); and (b-c) AuNPs@2H-MoS$_2$ hybrids synthesized in DMF by hot-injection of (b) gold precursor and citrate (*DC0*); and (c) gold precursor and double concentration of citrate (*D2C*).

- **Electrochemical Characterization**

Figure S8 represents the cyclic voltammetric (CV) studies, illustrating the electro-oxidation of Au in $H_2SO_4$ (Fig. S8a) and the $Fe(CN)_6^{3-/4-}$ redox activity (Figs. S8b-c), for pristine semiconducting 2H-MoS$_2$ (*M*) and the AuNPs@2H-MoS$_2$ hybrids (*A2M*).

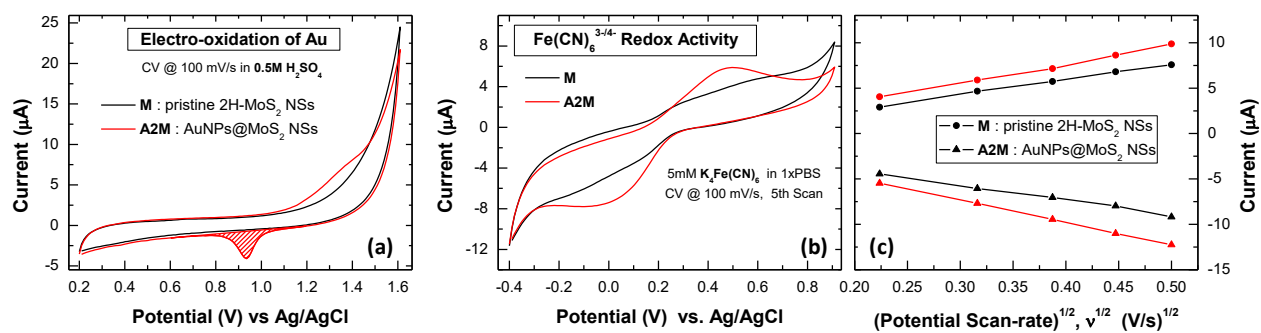

**Figure S8.** Electrochemical performances of pristine semiconducting 2H-MoS$_2$ NSs (*M*) and AuNPs@2H-MoS$_2$ hybrids synthesized in DMF by hot-injection of double concentration of gold precursor and citrate (*A2M*). (a) CV spectra recorded in 0.5 M $H_2SO_4$ at 100 mV/s vs. Ag/AgCl RE. (b) CV responses in 5 mM $K_4Fe(CN)_6$ (in 1x PBS) at 100 mV/s. (c) Redox peak currents as a function of scan rates for the $Fe(CN)_6^{3-/4-}$ redox activities shown in Figure S8b.

**Electrochemical Characterization of AuNPs.** The CV response of AuNPs@2H-MoS$_2$ hybrid (*A2M*) shown in Figure S8a is typical for gold electrodes in $H_2SO_4$ solution. The oxidation of AuNP-surface (a broad anodic peak appeared around 1.35 V) during the forward scan, is followed by subsequent reduction of the formed oxides during the backward scan (a well-defined reduction peak at 0.93 V), according to the following reaction:

$$Au + H_2O \leftrightarrow AuO + 2H^+ + 2e^- \qquad (1)$$

The electroactive surface area of AuNP can be represented by the charge ($Q_{oxides}$) integration under the oxide reduction peak (at 0.93V).[2-3] For *A2M* hybrids, the value of $Q_{oxides}$ is found to be 0.355 μC.

**Redox Activity of $Fe(CN)_6^{3-/4-}$ at AuNPs@2H-MoS$_2$.** The interfacial electro-catalytic properties of AuNPs@2H-MoS$_2$ hybrids were probed by the $Fe(CN)_6^{3-/4-}$ redox activity in Figure S8b. The AuNPs@2H-MoS$_2$ hybrids (*A2M*) exhibit obvious redox peak currents, relative to pristine semiconducting 2H-MoS$_2$ NSs (*M*), indicating a higher electroactive surface area and enhanced catalytic performance by providing improved conductivity and interconnectivity between the 2H-MoS$_2$ NSs.

Figure S8c reveals linear relationship of the redox peak currents ($I_p$) with the square root of scan-rates ($v^{1/2}$) of potential, for both the pristine 2H-MoS$_2$ (*M*) and AuNPs@2H-MoS$_2$ hybrids (*A2M*), suggesting diffusion-controlled mass transport, following the Randles-Sevcik equation:[3]

$$I_p = (2.69 \times 10^5) n^{3/2} \times A \times D^{1/2} \times C \times v^{1/2} \qquad (2)$$

where n is the number of electrons participating in the redox reaction. D and C represent the diffusion coefficient ($7.6 \times 10^{-6}$ cm$^2$/s) and concentration (mol/cm$^3$) of $K_4Fe(CN)_6$ in solution. A is the only parameter of the working electrode material, representing the electroactive surface area of electrode (cm$^2$). Hence $A_{Au@2H-MoS2}/A_{2H-MoS2}$ can define the enhancement of electrocatalytic performance of AuNPs@2H-MoS$_2$ hybrids.[3] The electroactive surface area of *A2M* was 1.24 times higher than that of pristine semiconducting 2H-MoS$_2$ (*M*).

**Hydrogen Evolution Reaction (HER) Study.** HER studies were performed in $N_2$-rich 0.5 M $H_2SO_4$ (aqueous) solution employing a rotating disc electrode (RDE) configuration. Initially, the WE was pre-conditioned by performing cyclic voltammetry for at least 20 scans, at a scan-rate of 100 mV/s. Polarization study was carried out by performing linear sweep voltammetry at a scan-rate of 5 mV/s. The measured polarization curves are reported by converting the potential from Ag/AgCl (3M KCl) scale to the reversible hydrogen electrode (RHE) scale, following the relation:

$$E_{RHE}^{(reported)} = E_{Ag/AgCl}^{(measured)} - (0.059 \times pH) - E_{Ag/AgCl}^0 \quad (3)$$

$$E_{Ag/AgCl}^0 = 0.21\ V\ and\ pH = 0\ for\ H_2SO_4$$

The measured polarization curves are reported after iR correction in order to compensate for any potential loss arising from external Ohmic resistance (R) of the electrochemical system. The R values were estimated by accounting the impedance value at the high frequencies ($10^5$ Hz), measured by electrochemical impedance spectroscopy (EIS). The EIS was performed over the frequency range from 1 MHz to 10 mHz, at a bias of -0.2 V (vs. RHE) superimposing a small sinusoidal voltage of 10 mV.

- **REFERENCES**